\documentclass[11pt]{article}
\usepackage{setspace,braket}
\usepackage{amsmath, amssymb, bm, yfonts}
\usepackage{etoolbox}
\usepackage{jheppub}
    \makeatletter
    \patchcmd{\maketitle}{\@fpheader}{}{}{}
\makeatother
\usepackage{booktabs}
\usepackage{multirow}
\usepackage{longtable}
\usepackage{tikz}
\usepackage{graphicx}
\def\be{\begin{equation}}
\def\ee{\end{equation}}

\def\be{\begin{equation}}
\def\ee{\end{equation}}

\def\bg{\bar{g}}

\def\beq{\begin{eqnarray}}\def\eeq{\end{eqnarray}}
\def\ba#1\ea{\begin{align}#1\end{align}}
\def\bg#1\eg{\begin{gather}#1\end{gather}}
\def\bm#1\em{\begin{multline}#1\end{multline}}
\def\bmd#1\emd{\begin{multlined}#1\end{multlined}}

\def\({\left(}
\def\){\right)}
\def\[{\left[}
\def\]{\right]}

\begin{document}

\title{A Framework for Non-linear Time Evolution and Field Theories on State-Dependent Geometries}
\author[a,b]{Dushyant Kumar}
\affiliation[a]{Harish-Chandra Research Institute, Chhatnag Road, Jhusi, Allahabad 211019, India}
\affiliation[b]{Homi Bhabha National Institute, Anushakti Nagar, Mumbai 400085, India}
\emailAdd{sehrawat.dushyant@gmail.com}
\begin{abstract}{ We introduce a framework for non-linear time evolution in quantum mechanics as a natural non-linear generalization of the Schrodinger equation. Within our framework, we derive simple toy models of dynamical geometry on finite graphs. Along similar lines we also propose a model of non-linear quantum field theory on spaces with state-dependent geometry. }
\end{abstract}
\maketitle
\onehalfspace

\section{Introduction}
It is well known for a long time that if we consider a quantum field theory on a curved background and consider the problem of computing back reaction of matter on gravity we get a non-linear quantum theory~\cite{Kibble1, Kibble2, BM, Dio, RP}. Naively one needs to deal with the following system of equations
\begin{equation}
\begin{aligned}[c]
i\frac{d|\psi\rangle}{dt} & = H(g)|\psi\rangle\\
G_{\mu\nu}(g) & =  8\pi \langle \psi| T_{\mu\nu}|\psi\rangle.\\
\end{aligned}
\end{equation}
Since both the equations have a dependence on the quantum state, the time evolution is effectively non-linear. In words of Mielnik~\cite{BM} ``{\it Either gravity is not classical, or quantum mechanics is not orthodox}". However, even if gravity is quantum and quantum mechanics is orthodox, non-linear quantum theory may still be useful in understanding some aspects of gravity. Non-linear quantum mechanics, in the form of state-dependent operators, has also registered its presence in the context of ADS-CFT in the description of interior of a black hole in Papadodimas-Raju proposal~\cite{PR}. This may possibly be an indication of a deeper role of non-linear quantum mechanics in the final formulation of quantum theory of gravity.

Several models for non-linear quantum theories have been proposed in the past~\cite{Kibble3, SW, MT}. In particular, frameworks for non-linear quantum theories proposed by Weinberg~\cite{SW} and Kibble~\cite{Kibble3} are general enough to accommodate various important special cases within them. The aim of the present work is to propose a new framework for non-linear time evolution. We then use this framework to generate simple toy models of dynamical geometry on finite graphs. In close analogy with the toy model of non-linear QFT on finite graphs, we also propose a model of non-linear quantum theory on continuum spaces with state-dependent geometry. We hope that such models of non-linear QFT may be useful as models somewhere intermediate between {\it QFT on a curved spacetime} and a {\it full quantum theory of gravity}.

This paper is organized as follows. We start with an introduction to our framework in Section~\ref{Intro}. This section is divided into two small subsections : In the first subsection ~\ref{GenIntro} we give a general introduction and discuss some properties of the framework, then in subsection~\ref{Examp} we give some examples. In section~\ref{Graphs} we use our framework in the context of quantum mechanics and field theory on finite graphs and derive toy models of dynamical geometry on them.  In Section~\ref{Continuum} we introduce the notion of state-dependent geometry in the continuum and propose a model of non-linear scalar field theory on such spaces. We end with conclusions in Section~\ref{Concl}.

\section{Introduction to the framework \label{Intro}}
\subsection{General aspects \label{GenIntro}}
The basic idea for our proposal comes from Random Matrix theory (RMT). In RMT one studies statistical spectral properties of an ensemble of Hamiltonians. We extend this idea into dynamical regime and propose time evolution under the average of a fixed ensemble of Hamiltonians where the probability distribution of the ensemble is taken to be state-dependent. More precisely, the idea is following. Instead of a single Hamiltonian consider an ensemble of Hamiltonians $\{H_1(\vec{l}_1),\dots,H_k({\vec{l_k}}),\dots\}$. When the system is in state $|\psi\rangle$, the probability $p_i$ associated with $H_i$ is taken to be of the form  
\begin{equation}
p_i =  \frac{1}{Z}\exp\left(-\beta \langle\psi|H_i|\psi\rangle - \alpha f(H_i;\vec{l}_i)\right)
\label{eq:ProbFun}
\end{equation}
where, $f(H_i;\vec{l}_i)$ is a real valued function of $H_i$ and (or) the parameters $\vec{l}_i$ appearing in $H_i$, $\alpha$ and $\beta$ are two real dimensionful constants, and
\begin{equation}
Z=\sum_{i}\exp\left(-\beta\langle\psi|H_i|\psi\rangle-\alpha f(H_i;\vec{l}_i)\right).
\label{eq:PartFun}
\end{equation}
Time evolution (of a normalized state) is now defined to be under the average Hamiltonian of the ensemble. That is,
\begin{equation}
i\frac{d|\psi\rangle}{dt}= H_{avg}|\psi\rangle
\label{Eq:TE}
\end{equation}
where, 
\begin{equation}
H_{avg} =\frac{1}{Z}\sum_{i}H_i\exp\left(-\beta\langle\psi|H_i|\psi\rangle-\alpha f(H_i;\vec{l}_i)\right).
\end{equation}
The formalism comes naturally equipped with its own thermodynamics where the thermodynamic quantities are defined by taking $Z$ as the partition function. In particular, one can define an energy functional 
\begin{eqnarray}
U = \langle\psi|H_{avg}|\psi\rangle & = & \frac{1}{Z}\sum_{i}\langle\psi|H_i|\psi\rangle\exp\left(-\beta\langle\psi|H_i|\psi\rangle-\alpha f(H_i;\vec{l}_i)\right)\nonumber\\
& = & -\displaystyle\frac{\partial}{\partial\beta}\bigg|_r\log(Z)
\end{eqnarray}
and the entropy functional 
\begin{eqnarray}
S=-\sum_i p_i\,\log(p_i)=-\left(\beta\displaystyle\frac{\partial}{\partial\beta}+\alpha\displaystyle\frac{\partial}{\partial\alpha}-1\right)\bigg|_r\log(Z)
\end{eqnarray}
where, $\big|_r$ in these expressions denotes the restricted partial derivative which neglects the implicit dependence of the state $|\psi\rangle$ on $\alpha$ and $\beta$. Of course, the notation $\big|_r$ makes sense only when $Z$ is written in the form \eqref{eq:PartFun}.  

It is easy to show that the partition function $Z$ is constant in time. In fact, denoting $\beta \langle\psi|H_j|\psi\rangle + \alpha f(H_j;\vec{l}_j)$ as $P(H_j)$, we have, 
\begin{eqnarray}
\frac{dZ}{dt}&=&-\beta\sum_{j}\exp\left(-P(H_{j})\right)\left(\left(\frac{d}{dt}|\psi\rangle,\: H_{j}|\psi\rangle\right)+\left(|\psi\rangle,\: H_{j}\frac{d}{dt}|\psi\rangle\right)\right)\nonumber \\ &=&-\beta\sum_{j}\exp\left(-P(H_{j})\right)\left(\left(-iH_{avg}|\psi\rangle,\: H_{j}|\psi\rangle\right)+\left(|\psi\rangle,-iH_{j}H_{avg}|\psi\rangle\right)\right)\nonumber \\&=&-i\beta\sum_{j}\exp\left(-P(H_{j})\right)\left(|\psi\rangle,[H_{avg},\: H_{j}]|\psi\rangle\right)\\&=&-i\beta\: Z\left(|\psi\rangle,\:[H_{avg},H_{avg}]|\psi\rangle\right)\nonumber\\&=&0. \nonumber
\end{eqnarray}
It is also clear that the time evolution introduced by Eq.~\eqref{Eq:TE} is norm-preserving. This follows from the fact that $H_{avg}$ is Hermitian. The time evolution equation can also be made scale invariant, i.e., invariant under $|\psi\rangle\to a|\psi\rangle,\,a\ne 0$, by using the following, slightly modified, probability function
\begin{equation}
p_i=\frac{1}{Z}\exp\left(-\beta \frac{\langle\psi|H_i|\psi\rangle}{\langle\psi|\psi\rangle}-\alpha f(H_i;\vec{l}_i)\right).
\label{eq:ProbFun2}
\end{equation}
However, since the norm is preserved, one can always use the initial condition $\langle\psi|\psi\rangle=1$ to write the probability function in the form given by Eq.~\eqref{eq:ProbFun}.

There are some conditions under which the time evolution defined by Eq.~\eqref{Eq:TE} reduces to the ordinary Schrodinger equation. If we have a degenerate ensemble, i.e., an ensemble containing a single Hamiltonian $H$, then $H_{avg}=H$ and hence, the time evolution becomes ordinary time evolution under $H$. One other situation when the time evolution becomes linear\footnote{By ``linear time evolution" we mean time evolution under ordinary Schrodinger equation in which the Hamiltonian does not have any dependence on the state.} is when the Hamiltonians of the ensemble pairwise commute with each other. To see this, note that,
\begin{eqnarray}
\frac{d}{dt}H_{avg}&=&-\frac{i\beta}{Z}\sum_{j}\exp\left(-P(H_{j})\right)\left(|\psi\rangle,[H_{avg},H_{j}]|\psi\rangle\right)H_{j}\nonumber\\&=&-\frac{i\beta}{Z^{2}}\sum_{j,k}\exp\left(-P(H_{j})-P(H_{k})\right)\langle\psi|[H_{k},H_{j}]|\psi\rangle H_{j}.
\end{eqnarray}
Therefore, if $[H_{k},H_{j}]=0$ for all $k,j$, then $\displaystyle\frac{d}{dt}H_{avg}=0$, and hence, the time evolution equation becomes linear. In particular, an ``almost commuting" ensemble of Hamiltonians is expected to give rise to an ``almost linear" time evolution. 
\subsection{Examples \label{Examp}}
In this section we consider two simple examples of non-linear evolution involving i) a discrete and ii) a continuum ensemble. For the discrete case, consider the ensemble containing just three matrices - the Pauli matrices \textcolor{black}{$\sigma_1, \sigma_2,\sigma_3$} with the probability function 
\begin{equation}
p(\sigma_i)=\frac{1}{Z}\exp\left(-\beta \langle\psi|\sigma_i|\psi\rangle\right).
\end{equation}
The equation of motion is written as
\begin{eqnarray}
i\frac{d|\psi\rangle}{dt} & = &\frac{1}{Z}\left(\exp\left(-\beta\langle\psi|\sigma_1|\psi\rangle \right) \sigma_1+\exp\left(-\beta\langle\psi|\sigma_2|\psi\rangle \right)\sigma_2+\right.\nonumber\\
{}&{}& \left. \exp\left(-\beta\langle\psi|\sigma_3|\psi\rangle \right)\sigma_3\right)|\psi\rangle.
\end{eqnarray} 
Since the Pauli matrices are pairwise strongly non-commuting, the time evolution is expected to be highly non-linear. This makes it difficult to find analytic solutions. However, the system can be evolved for small time intervals using numerical methods. Figure~\ref{fig:energy} shows a plot of expectation values of matrices $\sigma_1,\,\sigma_2,\,\sigma_3$ and $\sigma_{avg}$ in the state $|\psi(t)\rangle$ with time $t$. The average energy $\langle\psi|\sigma_{avg}|\psi\rangle$ changes much slowly compared to the expectation values of individual matrices $\sigma_1,\,\sigma_2,\,\sigma_3$ of the ensemble. We expect this to be a general feature of the time evolution in our framework.  
\newsavebox{\smlmat}
\savebox{\smlmat}{$\frac{1}{\sqrt{2}}\left[\begin{array}{c} 1 \\ 1 \end{array} \right]$}
 \begin{figure}[h!]
 \centering
  \includegraphics[width=0.80\textwidth]{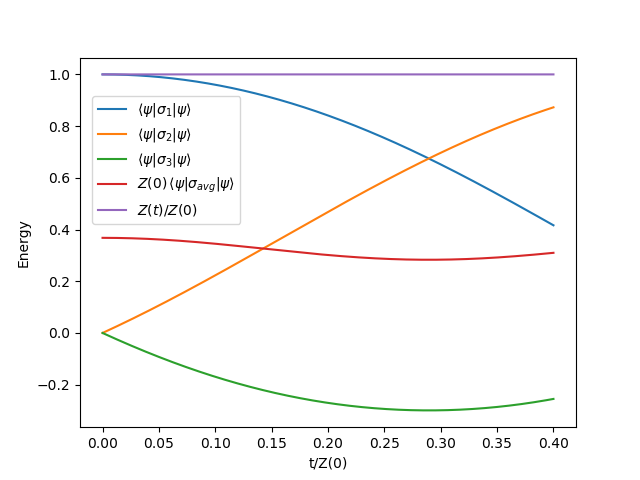}
 \caption{\small Plot of energy {\it vs} time for the sigma ensemble with initial state~\usebox{\smlmat} and $\beta=1$.}
 \label{fig:energy}
 \end{figure}

The non-linear equation tends to simplify for continuously infinite ensembles. For example, consider the ensemble of $N\times N$ real symmetric matrices, with the probability function
\begin{equation}
p(H)=\frac{1}{Z}\exp\left(-\beta\langle\psi|H|\psi\rangle-\alpha \text{Tr}(H^2)\right).
\label{ProbFun3}
\end{equation}
For $H=(a_{ij})$ and $|\psi\rangle =(x_1+iy_1,\dots,x_N+iy_N)^t$ we have,
\begin{equation}
\langle\psi|H|\psi\rangle=\sum_{i=1}^{N}a_{ii}(x_{i}^{2}+y_{i}^{2})+2\sum_{i=1}^{N}\sum_{j=i+1}^{N}a_{ij}(x_{i}x_{j}+y_{i}y_{j})
\end{equation}
and
\begin{equation}
\mbox{Tr}(H^{2})=\sum_{i=1}^{N}a_{ii}^{2}+2\sum_{i=1}^{N}\sum_{j=i+1}^{N}a_{ij}^{2}.
\end{equation}
The integration measure over the ensemble is taken to be
\begin{equation}
d\mu(H)=\displaystyle\prod_{i=1}^{N}da_{ii}\prod_{j>i}da_{ij}.
\end{equation}
Therefore, the average Hamiltonian for the ensemble of real symmetric matrices comes out to be $H_{avg}=(l_{ij})$ where,
\begin{eqnarray}
l_{ii}& = &-\frac{\beta}{2\alpha}(x_i^2+y_i^2),\nonumber\\
l_{ij} & = &-\frac{\beta}{2\alpha}(x_ix_j+y_iy_j)\,\,\text{for}\,\,i\ne j.
\end{eqnarray}
 This is nothing but the matrix $-\displaystyle \frac{\beta}{2\alpha}(|\psi\rangle \langle \psi|)_{real}$. Hence, the time evolution equation in this case admits a very simple form
\begin{eqnarray}
i\frac{d|\psi\rangle}{dt} = -\frac{\beta}{2\alpha}(|\psi\rangle \langle \psi|)_{real} |\psi\rangle.
\end{eqnarray}
If we instead consider the ensemble of all $N\times N$ Hermitian matrices with the same probability distribution as given by Eq.~\eqref{ProbFun3} then the average Hamiltonian comes out to be $-\displaystyle \frac{\beta}{2\alpha}|\psi\rangle \langle \psi|$. Therefore, in this case, the time evolution reduces to the ordinary Schrodinger equation and is given as 
\begin{eqnarray}
i\frac{d|\psi\rangle}{dt} = -\frac{\beta}{2\alpha}I_N |\psi\rangle
\end{eqnarray}
where, $I_N$ denotes the identity matrix of order $N$. 
\section{Non-linear theories on finite graphs\label{Graphs}}
\subsection{Notation}
A finite graph consists of a set $V=\{v_1,\dots,v_n\}$ of vertices and a set $E=\{e_s\mid s\in\{(i,j)\mid i,j=1,\dots,n\}\}$ of edges. We will consider only undirected simple graphs so that an edge $e_{ij}$ would be same as edge $e_{ji}$ and edges from a vertex to itself will not be allowed. If vertices $i,j$ are neighbouring vertices then we will denote this as $<i,j>$. 

We would also associate edges of a graph with positive real numbers called weights. These weights can be thought of as inverse length squares of the edges and, hence, define a geometry on the graph. We term a graph without any weights associated with edges as a topological graph, while a graph with weighted edges is termed as weighted or geometric graph. Consider a geometric graph with $N$ vertices labelled as $i=1,\dots,N$ and with weights $a_{ij}=a_{ji}$ associated with edges, where, $a_{ij}=0$ if $i,j$ are not neighbouring vertices or if $i=j$. We associate with any such graph a Laplacian matrix, also called Kirchhoff matrix, $\nabla^2=(d_{ij})$ where, $d_{ij}=a_{ij}$ for $i\ne j$ and $d_{ii}=-\displaystyle\sum_{j}a_{ij}$. The reason for calling this matrix Laplacian is simply because it satisfies the equation
\begin{eqnarray}
 \sum_{<i,j>}a_{ij}(\phi_i-\phi_j)^2 = - \phi \nabla^2 \phi\,\,\, \text{where}\,\,\,\phi=(\phi_1,\dots,\phi_N)^t
\end{eqnarray}
which is analog of the corresponding equation 
\begin{equation}
 \int d^n x\sqrt{g}\,g^{ij}\partial_{i}\phi\partial_j\phi = - \int d^n x\sqrt{g}\,\phi\nabla_{x}^2 \phi
\end{equation}
satisfied by the Laplacian operator in the continuum. 
\subsection{Ensemble of free particle Hamiltonians on a finite graph} 
The Hamiltonian $H$ of a free particle of mass $m$ on a graph with $N$ vertices labelled as $i=1,\dots, N$ and with Laplacian matrix $\nabla^2$ is given as $H=-\displaystyle\frac{1}{2m}\nabla^2$. Consider the ensemble of all these Hamiltonians for all different graph geometries on a graph of fixed topology. The probability function in a state $|\psi\rangle =(\psi_1,\dots,\psi_N)^t$ is taken to be
\begin{equation}
p(H)=\frac{1}{Z}\exp\left(-\beta\langle\psi|H|\psi\rangle+\frac{\alpha}{2}\text{Tr}(\nabla^2)\right)
\end{equation} 
and the integration measure is defined as,
\begin{equation}
d\mu=\prod_{<ij>}d a_{ij}
\label{Eq:measure2}
\end{equation}
where the range of integration for each edge weight is $(0,\infty)$. The average Hamiltonian is easily computed and comes out to be  $H_{avg}=(l_{ij})$, where,
\begin{align}
 l_{ij} &= 0           & \text{if} & \text{ $i\ne j$ and $i,j$ are not neighbors}\nonumber\\
 l_{ij} &= \frac{2m}{\beta |\psi_i-\psi_j|^2 + 2m\alpha}     & \text{if} & \text{ $i\ne j$ and $i,j$ are neighbors}\\
 l_{ii} & = -\sum_{j\ne i} l_{ij}.     &  \nonumber
\end{align}
Some important points to note from this toy example are 
      \begin{itemize}
      \item The average Hamiltonian corresponds to a particle on a graph with state-dependent geometry.
      \item At each instant of time, the quantum state contains full information about the geometry. As the state evolves in time, so does the geometry.  
      \item In the vacuum state $\frac{1}{\sqrt{N}}(1,\dots,1)^t$, all the edges are of same weight and the time evolution reduces to linear. 
        \end{itemize}
\subsection{Ensemble of scalar field theory Hamiltonians on a finite graph} 
As in the previous example, consider a geometric graph with $N$ vertices numbered as $i=1,\dots,N$ and with weights $a_{ij}=a_{ji}$ associated with edges, where, $a_{ij}=0$ if $i,j$ are not neighbouring vertices or if $i=j$. We will consider the case of a scalar field theory. Therefore, we attach quantum degrees of freedom $\hat{\pi}_i,\hat{\phi}_i$ to each vertex $i$. The Hamiltonian of the field theory would be of the form
 \begin{eqnarray}
 H &=&\frac{1}{2}\sum_{j=1}^N\hat{\pi}_j^2+\frac{1}{2}\sum_{<ij>}a_{ij}(\hat{\phi}_{i}-\hat{\phi}_{j})^{2}
 +\frac{1}{2}m^2\sum_{i=1}^N\phi_i^2 + \frac{\lambda}{n!}\sum_{i=1}^N\phi_i^n.
\end{eqnarray}
Now, consider the ensemble of all these QFT Hamiltonians for all different graph geometries on a graph of fixed topology. We take our probability function in a state $|\psi\rangle$ to be 
\begin{equation}
p(H)=\frac{1}{Z}\exp\left(-\beta\langle\psi|H|\psi\rangle+\frac{\alpha}{2}\text{Tr}(\nabla^2)\right).
\end{equation}
The integration measure remains same as defined in Eq.~\eqref{Eq:measure2}. The average Hamiltonian is again easy to compute and comes out to be
\begin{eqnarray}
 H_{avg} &=&\frac{1}{2}\sum_{j=1}^N\hat{\pi}_j^2+\frac{1}{2}\sum_{<ij>}l_{ij}(\hat{\phi}_{i}-\hat{\phi}_{j})^{2}
+\frac{1}{2}m^2\sum_{i=1}^N\phi_i^2 + \frac{\lambda}{n!}\sum_{i=1}^N\phi_i^n
\end{eqnarray}
where, for neighbouring $i,j$
\begin{equation}
l_{ij} = \frac{1}{\frac{\beta}{2}\langle\psi|(\hat{\phi}_i-\hat{\phi}_j)^2|\psi\rangle + \alpha}.
\end{equation}
As in the example of the ensemble of Hamiltonians of a free particle considered before, the only effect of the ensemble averaging is to make the geometry of the graph state-dependent. The average Hamiltonian is highly non-linear but it is possible to compute the vacuum solution under some special conditions. For example, the vacuum state of a free theory on a graph with discrete translation invariance can be found in the following steps. 
         \begin{itemize}
         \item Find the vacuum state of the linear theory on the graph where each edge is assigned a fixed weight $a^2$. Lets denote this vacuum state as $|0,a^2\rangle$.
        \item Compute $l_{ij}=\displaystyle\frac{1}{\frac{\beta}{2}\langle 0,a^2|(\hat{\phi}_i-\hat{\phi}_j)^2|0,a^2\rangle + \alpha}$ for any fixed neighbouring sites $i,j$. The result will be independent of the choice of $i$ and $j$ due to translational invariance. Denote the function so obtained as $g(a^2,\beta,\alpha)$. 
        \item Solve the equation $g(a^2,\beta,\alpha)=a^2$ for $a$ in terms of $\alpha$ and $\beta$. Let the solution be $a_0$.
        \item Then the required vacuum state will be $|0,a_0^2\rangle$.
         \end{itemize}
The time evolution of the vacuum state is governed by the linear Schrodinger equation. Therefore, any state which deviates only a little from the vacuum state is expected to follow almost linear time evolution.      
\section{QFT on continuum spaces with state-dependent geometry\label{Continuum}}
It is difficult to apply our formalism of non-linear time evolution to continuum spaces simply because the moduli spaces of metrics in the continuum are infinite dimensional and integration over them is highly non-trivial. However, one can try to formulate models of non-linear QFT on continuum spaces in analogy with the toy model of non-linear QFT on graphs. The only condition we impose on such models is that the geometry be state-dependent and reduce to flat geometry, with linear time evolution for the vacuum state. The most important information one needs for defining a field theory on a state-dependent geometry is an expression for the metric in terms of the expectation value of a local operator. The geometry in the case of a finite graph depends on the state through the expectation value $\langle\psi|(\hat{\phi}_i-\hat{\phi}_j)^2|\psi\rangle $. In the continuum limit this quantity would be similar to $\langle\psi|\left(\partial_x\hat{\phi}(x)\right)^2|\psi\rangle \sim \displaystyle\lim_{x\to y}\partial_x \partial_y\langle\psi|\hat{\phi}(x)\hat{\phi}(y)|\psi\rangle $. Therefore, we expect that the metric in case of continuum spaces should somehow be recoverable from the singular limit of the double derivatives of the Green's function. Such an expression for the the metric in terms of Green's function has already been derived in~\cite{SAK}
\begin{equation}
g_{ij}(x)=-\frac{1}{2}\left(\frac{\Gamma\left(\frac{D-2}{2}\right)}{4\pi^{D/2}}\right)^{\frac{2}{D-2}} \lim_{x\to y}\frac{\partial}{\partial x^i}\frac{\partial}{\partial y^j}\left(G(x,y)^{\frac{2}{2-D}}\right).
\label{metric}
\end{equation}
Here $D$ is the dimension of spacetime which is required to be $\ge 3$ for the above result to be valid.

Using the result~\eqref{metric} one can define a non-linear scalar field theory on a space of the form $\mathbb{R}^d\times\mathbb{R}$ with a state-dependent metric of the form $dt\otimes dt+h_{ij}(x,t)dx^i\otimes dx^j$ as follows. 
First we choose a spatial slice $\mathbb{R}^d\times\{t_0\}$ and attach quantum degrees of freedom $\hat{\pi}(t_0,\vec{x}),\,\hat{\phi}(t_0,\vec{x})$ to each point $(t_0,\vec{x})$ of this slice. These degrees of freedom are required to satisfy the commutation relation $[\hat{\phi}(\vec{x}),\hat{\pi}(\vec{y})]=i\delta^d(\vec{x}-\vec{y})$ where $\delta^d(\vec{x}-\vec{y})$ is the delta function on $\mathbb{R}^d\times\{t_0\}$ with respect to a flat metric. We can now define the state-dependent metric on the spatial slice as 
\begin{equation}
h_{ij}(\vec{x})=-\frac{1}{2}\left(\frac{\Gamma\left(\frac{d-1}{2}\right)}{4\pi^{(d+1)/2}}\right)^{\frac{2}{d-1}} \lim_{\vec{x}\to\vec{y}}\frac{\partial}{\partial x^i}\frac{\partial}{\partial y^j}\left(\langle\psi|\hat{\phi}(\vec{x})\hat{\phi}(\vec{y})|\psi\rangle^{\frac{2}{1-d}}\right)
\end{equation}
Using this state-dependent metric the Hamiltonian for the non-linear scalar field theory can be written as
\begin{equation}
H=\frac{1}{2}\int d^d x \,\left(\frac{\hat{\pi}^2(\vec{x})}{\sqrt{h}}+\sqrt{h}\,h^{ij}\partial_i\hat{\phi}\,\partial_j\hat{\phi}+m^2\sqrt{h}\,\hat{\phi}^2\right)
\end{equation}
The non-linear field theory corresponding to the above Hamiltonian has some level of similarity with the non-linear field theory on finite graphs considered before. The vacuum state is simply the flat space vacuum of the scalar field. Also, the theory would reduce to a linear theory if the initial state is the vacuum state, and hence, can be expected to be ``close to linear" for any state ``close to vacuum". However, in contrast to the theory on finite graphs, we are not yet sure if this theory can somehow be derived within our formalism. 
\section{Conclusions\label{Concl}}
We have introduced a framework for non-linear time evolution by replacing the idea of ``{\it time evolution under a fixed Hamiltonian}" with ``{\it time evolution under an ensemble of Hamiltonians with state-dependent probability distribution}". The framework naturally comes equipped with its own thermodynamics. The physical meaning and dynamical behaviour of quantities like entropy is yet to be understood. 

Within our framework, we have also derived simple toy models of dynamical geometry on finite graphs. Moreover, in close analogy with these toy models, we have introduced a model of non-linear scalar field theory on continuum spaces with state-dependent geometry. An important aspect of these models is that the quantum state of matter contains full information about the spatial geometry. Hence, these non-linear models provide simplest examples in which the geometry evolves with the quantum state. We leave it to future work to study the relevance of these models to questions in quantum gravity.


\end{document}